# Widely Tunable Berry curvature in the Magnetic Semimetal Cr$_{1+\delta}$Te$_2$

Y. Fujisawa[1], M. Pardo-Almanza[1], C. H. Hsu[1,2,3], A. Mohamed[1], K. Yamagami[1], A. Krishnadas[1],

F. C. Chuang[2,3,4], K. H. Khoo[5], J. Zang[6,7], A. Soumyanarayanan[8,9], Y. Okada[1,]

[1]*Quantum Materials Science Unit, Okinawa Institute of Science and Technology (OIST), Okinawa 904-0495, Japan*

[2] *Department of Physics, National Sun Yat-sen University, Kaohsiung 80424, Taiwan*

[3]*Physics Division, National Center for Theoretical Sciences, Taipei 10617, Taiwan.*

[4]*Department of Physics, National Tsing Hua University, Hsinchu 30013 Taiwan.*

[5] *Institute of High Performance Computing,*

*Agency for Science Technology and Research, 138632 Singapore*

[6] *Department of Physics and Astronomy, University of New Hampshire, Durham, NH 03824, USA.*

[7] *Materials Science Program, University of New Hampshire, Durham, NH 03824, USA.*

[8]*Department of Physics, National University of Singapore, 117551 Singapore*

[9] *Institute of Materials Research and Engineering,*

*Agency for Science Technology and Research, 138634 Singapore*

**Abstract**

Magnetic semimetals have increasingly emerged as lucrative platforms hosting spin-based topological phenomena in real- and momentum spaces. Cr$_{1+\delta}$Te$_2$ is a self-intercalated magnetic transition metal dichalcogenide (TMD), which exhibits topological magnetism, tunable electron filling, and magnetic frustration etc. While recent studies have explored real-space Berry curvature effects in this material, similar considerations of momentum-space Berry curvature have been lacking. Here, we systematically investigate the electronic structure and transport properties of epitaxial Cr$_{1+\delta}$Te$_2$ thin films over a wide range of doping, δ (0.33 – 0.71). Spectroscopic experiments reveal the presence of a characteristic semi-metallic band region near the Brillouin Zone edge, which shows a rigid band like energy shift as a function of δ. Transport experiments show that the intrinsic component of the anomalous Hall effect (AHE) is sizable, and undergoes a sign flip across δ. Finally, density functional theory calculations establish a causal link between the observed doping evolution of the band structure and AHE: the AHE sign flip is shown to emerge from the sign change of the Berry curvature, as the semi-metallic band region crosses the Fermi energy. Our findings underscore the increasing relevance of momentum-space Berry curvature in magnetic TMDs and provide a unique platform for intertwining topological physics in real and momentum spaces.

**Main Text**

**Introduction**

In recent years, fascinating new condensed matter phenomena have been found to arise from the interplay of spin degree of freedom of electrons with emergent geometric and topological effects [1,2]. Prominent among these is the concept of Berry curvature Ω, which arises from the geometric phase accumulated by electronic wavepackets traversing a closed loop [3,4]. In crystalline solids, such Berry curvature can be interpreted as an effective magnetic field acting on moving electrons, and therefore manifests prominently in Hall transport experiments [1]. For example, the quantization of its integral in the momentum space, a phenomenon known as band topology, results in quantized charge and spin Hall effects [5,6,7,8]. On the other hand, magnetic materials exhibit rich manifestations of real- and momentum-space Berry curvature [9,10]. In particular, noncollinear and chiral spin structures give rise to real space Berry curvature, which results in the topological Hall effect (THE) for itinerant electrons [10,11,12,13]. In contrast, the momentum space Berry curvature of Bloch wave functions of electronic bands contributes prominently to the anomalous Hall effect (AHE) [9,14,15,16], which may result in its quantization and unconventional non-linear magnetic field dependence [17,18,19,20,21]. The robustness of such k-space Berry curvature, its material tunability, and direct accessibility from experimental probes makes it attractive for electronic and optical applications [22].

Magnetic semimetals have served as attractive platforms in the search for Berry curvature driven phenomena, with the potential presence of Dirac and Weyl band physics [23,24,25,26,27,28,29,30]. Notably, the ability to realize systematic tuning of electron filling and magnetic properties in such materials without significantly altering the crystal structure is particularly attractive [31]. Recent works have focused on transition metal dichalcogenides (TMDs) due to the flexibility in incorporating intercalated atoms between weakly van der Waals coupled atomic layers [32,33,34], which enables tunability of magnetic properties. Interest in magnetic TMDs has grown rapidly due to their tunable magnetic properties and potential for hosting non-collinear spin textures [32,33]. In contrast, investigations of $k$-space Berry curvature in such TMDs has yet to receive significant attention due to the lack of reliable means to systematically control band structure with doping. The potential for realizing concomitant real- and momentum space Berry curvature in such systems is a promising prospect for the exploration of exotic topological phenomena [35,36].

Prominent among magnetic TMDs is $Cr_{1+\delta}Te_2$ – a self-intercalated TMD with tunable valency ($Cr^{2+}$ ($3d^4$) to $Cr^{4+}$($3d^2$), assuming simple ionic crystal picture) – which therefore serves as a unique candidate for modulating electronic and magnetic properties over a wide range (**Fig. 1A**) [37,38,39,40,41,42,43,44,45,46,47,48,49,50,51,52]. Our recent efforts have enabled a unique recipe to realize epitaxial thin films of $Cr_{1+\delta}Te_2$ over wide range of doping, with tunable magnetic properties [53]. Here, by combining in-situ angle resolved photoemission spectroscopy (ARPES), transport measurements, and density functional theory (DFT) calculations, we show that such $Cr_{1+\delta}Te_2$ films form a unique magnetic and semi-metallic platform that hosts widely tunable Berry curvature near the chemical potential in the momentum space.

## Results & Discussion

**Basic Characteristics of Cr-Te films**

In this work, epitaxial $Cr_{1+\delta}Te_2$ thin films were grown by molecular beam epitaxy (MBE) using an established recipe [53], which uniquely enables the doping level δ to be varied over a wide range (0.3 < δ < 0.8). We have previously shown that $Cr_{1+\delta}Te_2$ thin films maintain the NiAs-type crystal structure of $CrTe_2$ (**Fig. 1A**), while dopants $Cr_\delta$ intercalate between the weakly coupled Te layers [53]. Note that regardless of our extensive efforts, growing single crystalline phase with δ < 0.3 and δ > 0.8 was not possible [53]. **Figure 1B** shows the corresponding evolution of Curie temperature, $T_C(\delta)$, estimated from magnetometry and a characteristic kink in resistivity measurements (see ***SI appendix*, fig. S1**). Consistent with our previous report [53], $T_C(\delta)$ increases from 160 K, and saturates to ~ 350 K for $\delta \gtrsim 0.5$.

**Cr-Te Band Schematic**

We turn to examine the expected effect of doping on the band structure of $Cr_{1+\delta}Te_2$, whose bulk (lower) and surface (upper) Brillouin zone are shown in **Fig. 1C**. The schematic low energy band structure of $Cr_{1+\delta}Te_2$ – which is central to the key claims of this report – derived from ARPES experiments and DFT calculations is illustrated in **Fig. 1D** (detailed in subsequent sections). Near $\bar{\Gamma}$, there are three hole bands, which are denoted as $HB_{1\Gamma}$, $HB_{2\Gamma}$, and $HB_{3\Gamma}$ for convenience. Meanwhile, near $\bar{M}$, there are two hole bands ($HB_{1M}$, $HB_{2M}$) and one electron band ($EB_M$). With increasing number of intercalated cations, the Fermi energy, $E_F$, is expected to correspondingly shift upwards. The wide range of tunable electron filling in $Cr_{1+\delta}Te_2$, coupled with the semi-metallic character [54,55], motivates us to explore the evolution of its Berry curvature in momentum space (**Fig. 1C** and **D**), and its signature in band structure and transport measurements.

**ARPES Measurements**

The ARPES measurements are performed *in situ*, using an ultrahigh vacuum (UHV) cluster system that also connects the MBE system (**Fig. 2A**). The ARPES measurements are used to establish the semi-metallic nature of the near $E_F$ band structure of $Cr_{1+\delta}Te_2$ around the $\bar{M}$-point. Here, a semi-metallic band is characterized by an electron-like band branch at higher energies, and a hole-like band branch at lower energies coexisting and overlapping around the same momenta (see **Fig. 1D**, dashed circle at $\bar{M}$). For the illustrative purpose of **Fig. 2B-F**, we choose the $\delta \simeq 0.6$ sample as its higher electron filling provides ARPES – sensitive to occupied states – with wider energy access to the $Cr_{1+\delta}Te_2$ band structure.

**Fig. 2B** presents the $k$-resolved map of the Fermi surface, with clearly observable Fermi pockets at the $\bar{\Gamma}$ and $\bar{M}$ points. To further understand the character of these Fermi pockets, energy dependent ARPES spectral weights measured across $\overline{\Gamma}$ and $\overline{M}$ are shown in **Fig. 2C-D** and **Fig. 2E-F**, respectively. At $\overline{\Gamma}$ (**Fig. 2C-D**), we observe three hole bands ($HB_{1\Gamma}, HB_{2\Gamma}, HB_{3\Gamma}$), with the topmost –$HB_{1\Gamma}$– crossing $E_F$, and producing a hole-like Fermi pocket (**Fig. 2B**, right). Meanwhile, at $\overline{M}$ (**Fig. 2E-F**), we find an electron band ($EB_M$) crossing $E_F$ forms an electron-like Fermi pocket (**Fig. 2B**). Notably, in addition to this ($EB_M$), two hole bands ($HB_{1M}$ and $HB_{2M}$) exist slightly below $E_F$ at $\overline{M}$. Coexistence of electron and hole band branches at the $\overline{M}$ point hosts the near $E_F$ semi-metallic character in $Cr_{1+\delta}Te_2$.

Next, we shown the doping evolution of the band structure by studying ARPES spectral intensities, along $\overline{K} - \overline{\Gamma} - \overline{K}$ and $\overline{K} - \overline{M} - \overline{K}$ for samples with $\delta$ = 0.33, 0.4, 0.5, and 0.6. Consistent with the expected increase of electron filling with the increase of Cr cations, all bands at $\overline{\Gamma}$ and $\overline{M}$ are found to collectively shift downwards relative to $E_F$ with increasing δ (**Fig. 3A-H**). For $\delta < 0.5$ samples, the only occupied band – $HB_{2M}$ – is barely visible (**Fig. 3E-F**). In contrast, for $\delta \geq 0.5$, both the bottom of the electron band $EB_M$ and top of hole band $HB_{1M}$ are clearly seen below $E_F$, in addition to the lower hole band ($HB_{2M}$) (**Fig. 3G-H**). While intensity for HB$_{1M}$ between samples with $\delta$ = 0.5 (**Fig. 3G**) and $\delta$ = 0.6 (**Fig. 3H**) are largely different, weakly seen HB$_{1M}$ in sample with $\delta$ = 0.5 (**Fig. 3G**) is still in line with rigid-like band shift picture. Observation of rigid-like band shift is none trivial phenomenological finding, which capture essential doping evolution of electronic state for sample with 0.3 < $\delta$ < 0.8. As shown later, this finding plays a vital role to interpreting Hall effects.

To quantify the shift in band structure with doping, we further examine the evolution of various band features at $\overline{\Gamma}$ and $\overline{M}$. Notably, we find that, for an appropriate choice of δ-dependent $E_F$ shift, the dispersion of the measured bands across samples with varying δ (**Fig.3A-H**: markers) can be all collapsed onto a single dispersion, e.g. that of $\delta = 0.6$ sample (**Fig.3D,H**). The collapse of band dispersions for varying δ around $\overline{\Gamma}$ and $\overline{M}$ is summarized in **Fig. 3I** and **J**, respectively. The rigid-like shifts of energy bands thus observed at $\overline{\Gamma}$ and $\overline{M}$ are defined as $\Delta E_\Gamma$ and $\Delta E_M$, respectively, and their evolution with δ are plotted in **Fig. 3K**. The observed linear dependence of $\Delta E_\Gamma$ and $\Delta E_M$ on δ is consistent with the expected evolution of a quasi-2D band structure with parabolic dispersion. Linear fitting of the $\Delta E$ vs δ data in **Fig. 3K** yields a slope of 1.54 eV energy shift per intercalated Cr atom, i.e. $\Delta E_F / \Delta \delta \simeq 1.54$ eV/atom. This relation is consistent with DFT calculation (see **SI Appendix fig. S2**) and is especially useful for subsequent comparisons of experimental and theoretical Hall conductivities in the latter part of this report.

**Anomalous Hall Effect Measurements**

To visualize the manifestation of Berry curvature effects arising from the semi-metallic band component, we investigated the Hall transport properties of $Cr_{1+\delta}Te_2$ films, and their evolution with doping ($\delta$ = 0.33, 0.47, 0.51, 0.7), magnetic field ($B$), and temperature ($T$). For a conventional magnetic material with magnetization $M$, the Hall resistivity is phenomenologically given as

$$\rho_{yx}(B) = \rho_{yx}^O(B) + \rho_{yx}^A(M(B)).$$

Here, $\rho_{yx}^O$ and $\rho_{yx}^A$ are the ordinary and anomalous Hall resistivities respectively, wherein the latter incorporates effects of $k$-space Berry curvature [9]. Firstly, $\rho_{yx}^O(B) \equiv R_H B$, where $R_H$ is the Hall coefficient, can be straightforwardly determined by a linear fit to $\rho_{yx}$ in the high field limit, $B \gg B_S$, where $B_S$ is the out-of-plane (OP) saturation field estimated from magnetometry measurements (see *SI Appendix* fig. S3 for raw data). The AHE resistivity $\rho_{yx}^A(B)$ obtained upon subtracting $\rho_{yx}^O$ from $\rho_{yx}$ is plotted in **Fig. 4A-C** for a series of temperatures ($T/T_C \simeq$ 0.9, 0.5, $T = 2$ K) and doping. While the resulting AHE resistivity $\rho_{yx}^A(B)$ is typically proportional to the magnetization *M(B)*, the presence of chiral spin textures under intermediate fields may also induce a "topological" component that is non-monotonic in magnetization *M(B)* [10,11,12,13]. To avoid such complications, we focus on the saturated magnitude of AHE resistivity by determining its value at $B \simeq 40$ kOe $\gg B_S$. The saturated AHE resistivity, defined henceforth as $\rho_{yx}^{A_{sat}} \equiv \rho_{yx}^A(B > B_s)$, is indicated by horizontal dashed lines in **Fig. 4A-C**, and examined in detail subsequently.

As shown in **Fig. 4D-E**, the AHE ($\rho_{yx}^{A_{sat}}$) and the ordinary Hall coefficient ($R_H$) show clear differences in their evolution with temperature and doping. As in **Fig. 4D**, $R_H(T, \delta)$ persistently presents a positive sign. While the quantitative interpretation of $R_H$ is involved within multi-component Fermi surfaces [56], the persistence of a positive $R_H(T, \delta)$ indicates the dominant contribution of hole pockets at $\overline{\Gamma}$ or $\overline{M}$. Meanwhile, $\rho_{yx}^{A_{sat}}(T, \delta)$, in **Fig. 4E**, exhibits multiple sign changes – both across doping and temperature. For example, with decreasing temperature, for δ = 0.33, $\rho_{yx}^{A_{sat}}$ changes from positive to negative, while for δ = 0.7 it changes from negative to positive. Therefore, the sign of $\rho_{yx}^{A_{sat}}$ is likely unrelated to $R_H(T, \delta)$, and governed by a rich interplay of doping and temperature, which is indicative of corresponding variations in its constituent extrinsic and intrinsic AHE components [9,57].

To separate extrinsic and intrinsic AHE, a widely accepted starting point is to consider the total anomalous Hall conductivity (AHC) $\sigma_{xy}^A$ as arising from the linear combination of three independent conductivity channels: $\sigma_{xy}^A = \sigma_{sk} + \sigma_{sj} + \sigma_{int}$. Here, $\sigma_{sk}$ and $\sigma_{sj}$, are extrinsic skew and side-jump scattering terms, while $\sigma_{int}$ is the intrinsic Karplus-Luttinger term [9]. Next, we note with respect to the established scaling relation ($\sigma_{xy}^A$ vs. $\sigma_{xx}$) used to delineate these components that our sample is far from the region where intrinsic AHC dominates over extrinsic AHC components (see *SI appendix* **Fig. S4**) [9]. Thus, to separate intrinsic and extrinsic AHC components for our samples, we follow a phenomenological approach for disordered systems, adopted in several previous works [25,57,58,59,60]. For disordered systems, wherein extrinsic scattering is dominated by *T*-independent impurity scattering relative to *T*-dependent phonon scattering, the following modified AHC scaling relation has been proposed [57]

$$\sigma_{xy}^{A_{sat}} = (\alpha \sigma_{xx0}^{-1} + \beta \sigma_{xx0}^{-2}) \sigma_{xx}^2 + b \simeq \text{const.} \times \sigma_{xx}^2 + b. \quad (1)$$

Here, α, β and *b* represent contributions from skew scattering, side-jump, and intrinsic Karplus-Luttinger (Berry phase) derived terms, respectively. Notably, the extrinsic terms are scaled by the residual conductivity *σ*<sub>xx0</sub>, which relates to the residual resistivity $\rho_{xx0}$ arising from impurity-induced elastic scattering at the $T = 0$. Therefore, the intrinsic component ($b$) can be determined as the intercept of the plot of $\sigma_{xy}^{\text{Asat}}$ vs. $\sigma_{xx}^2$, as previously shown for various material systems [25,61,62,63].

Following the modified scaling relation (**Eq. 1**), the intercept $b$ ($\equiv \sigma_{xy}^{\text{expt}}$) extracted from fits to the linear component of $\sigma_{xy}^{\text{Asat}}$ vs. $\sigma_{xx}^2$ plots, shown in **Fig. 5A** for the studied samples, were used to estimate the respective intrinsic AHCs. In this work, to transform from resistivity to conductivity, we use the relations $\sigma_{xy}^{\text{Asat}} = \rho_{yx}^{\text{Asat}}/\rho_{xx}^2$, $\sigma_{xx0} = 1/\rho_{xx0}$, and $\sigma_{xx} = 1/\rho_{xx}$, given the small (<1%) contribution of transverse channel relative to the longitudinal channel. Note that the reliability of the modified scaling analysis is higher. Thus, to estimate intercept $b$ from $\sigma_{xy}^{\text{Asat}}$ vs. $\sigma_{xx}^2$ plots, we focus on region with lower $\sigma_{xx}$ magnitude, in particular for δ > 0.4 (see **Fig. 5A**). **Fig. 5B** shows thus obtained oping dependent $\sigma_{xy}^{\text{expt}}$ (left axis), which is contrasted with the corresponding variation of temperature averaged $\langle R_H \rangle$ (right axis). Crucially, the sign of $\sigma_{xy}^{\text{expt}}$ changes from positive to negative over δ ~ 0.33 – 0.47, while the sign of $\langle R_H \rangle$ remains persistently positive across doping. The expected origin of the measured sign change in $\sigma_{xy}^{\text{expt}}(\delta)$ is from the *k*-space Berry curvature evolution with doping (see **Fig. 5C**), and should be captured within band structure calculations.

**Comparison between Theory and Experiments**

We turn to *ab initio* density functional theory (DFT) calculations of the parent $CrTe_2$. To avoid modeling complicates of intercalated Cr atoms, we select DFT calculation regarding parent structure $CrTe_2$. This approach is also reasonable due to experimental observation of righd-like band shift (see **Fig. 3**). Regardless of this simplified approach, as we show hereafter, our sign change of AHE can be essentially interpreted by filling change on parent electronic state of parent $CrTe_2$ based on rigid-like band picture. To present this, we first need to match the energy level between ARPES measurements and DFT calculations. **Fig. 6A-B** shows the DFT band structure for $CrTe_2$ with varying $k_z$ overlaid on the ARPES data for δ ~ 0.6. First, the DFT calculations match the key qualitative features seen by ARPES, e.g., the electron and hole bands at $\overline{\Gamma}$ and $\overline{M}$, and notably the semi-metallic band feature at $\overline{M}$. Second, the DFT and ARPES band structures match quantitatively – with particular focus on the bottom of the electron band at $\overline{M}$ (see arrow in **Fig. 6B**) – when the measured $E_F$ for δ = 0.6 for ARPES is set to ~ 1.9 eV higher than the calculated $E - E_F(\text{calc.})$ for DFT (compared for all dopings in *SI Appendix* fig. S5). Such a shift between ARPES and DFT is consistent in sign with the expected filling of electrons with increasing δ, and in magnitude with the reported trends towards an underestimation of $E - E_F(\text{calc.})$ in *ab initio* calculations of this system [47] (see *SI Appendix*, **Section A** for validation of this reasonable shift from the view point of valency of Cr).

Since our ARPES results consistently evidence a rigid-like shift of the $Cr_{1+\delta}Te_2$ band structure with increasing δ, we compare $\sigma_{xy}^{\text{expt}}(\delta)$ with the theoretical value of $\sigma_{xy}^{\text{calc}}(E)$ obtained for the parent $CrTe_2$. Here, $\sigma_{xy}^{\text{calc}}(E)$ is calculated by integrating the total Berry curvature $\Omega_{xy}(E, \mathbf{k})$ across the whole Brillouin zone (see method section and **Eq. 2-3** therein). Notably, the one-to-one comparison between $\sigma_{xy}^{\text{calc}}(E)$ and $\sigma_{xy}^{\text{expt}}(\delta)$, shown in **Fig. 6C**, is enabled by combining the $E_F$ reference point between ARPES and DFT (**Fig. 6A-B**) with the measured $\Delta E_F/\Delta \delta$ (**Fig. 3K**). We observe excellent qualitative agreement between the transport and DFT results across δ. In particular, the calculations reproduces the sign change from negative to positive around δ ~ 0.5 (**Fig. 6C**: shaded area). Meanwhile, the quantitative suppression of the absolute value of $\sigma_{xy}^{\text{expt}}$ relative to $\sigma_{xy}^{\text{calc}}$ is presumably consistent with the expected reduction of intrinsic AHE in the disordered regime [9,19].

Importantly, the calculations further provide valuable insights on the $k$-space origin of the AHE sign change observed in DFT and transport experiments. In particular, understanding energy evolution of Berry curvature between 1.5 eV and 1.9 eV is essential due to sign flip of $\sigma_{xy}^{\text{calc}}(E)$ (see horizontal arrows and shaded area in **Fig. 6C**). We first examine in **Fig. 6D** the $k_z$-dependence of band dispersion along $\overline{K} - \overline{M} - \overline{K}$, which exhibits – in each case – a characteristic nearly degenerated band region with a spin-orbit coupling (SOC)-driven gap (dashed circles). Notably, this band region disperses substantially with $k_z$, consequently widening the energy window hosting prominent Berry curvature distribution (to ~0.2 eV). Next, the role of this band region in the sign change of $\sigma_{xy}^{\text{expt}}$ is elucidated by examing the $k_z$-integrated total Berry curvature, $\int dk_z \Omega_{xy}(E, \mathbf{k})$, shown in **Fig. 6E-F** for two different energies $E - E_F$(calc) = 1.5 eV (c.f. expt. δ ~ 0.3) and 1.9 eV (c.f. expt. δ ~ 0.6) on either side of the sign change (**Fig. 6C**: horizontal arrows). Notably for 1.9 eV, we observe, around the $\overline{M}$ point (**Fig. 6F**: vertical arrows), a prominent region hosting large positive $\int dk_z \Omega_{xy}(E, \mathbf{k})$, with momenta consistent with that hosting the SOC-driven gap (**Fig. 6D**: dashed circle). Contrastingly, for 1.5 eV (**Fig. 6E**), this particular source of total Berry curvature is absent around $\overline{M}$ point. Therefore, from **Eqn. 4** in method section ($\sigma_{xy}^{\text{calc}}(E) \propto -\int dk\Omega_{xy}(E, \mathbf{k})$), we can conclude that $\overline{M}$ point prominent positive Berry curvature at 1.9 eV is the leading cause of the sign change of $\sigma_{xy}^{\text{expt}}$ from positive to negative with increasing δ. Here, Weyl points (prominent sources of Berry curvature in other works [19,24]) – are also found in our DFT calculations for CrTe$_2$ (see ***SI Appendix* Section B**). However, their energies ($E - E_F$(calc) > 2 eV), are well above $E_F$ even for the highest doped sample. Thus, we can ignore the role of Weyl physics in our AHE results.

As an aside, the observed consistency of band structure between spin-resolved DFT calculations and ARPES experiments also provides a viable explanation for the saturation of $T_C$ with increasing doping noted here (**Fig. 1C**) and in previous works [53]. As shown in **Fig. 6D**, the upper branch of the semi-metallic band region, which hosts large Berry curvature, derives from minority spins (blue curve), and becomes occupied above δ ~ 0.5. Since $T_C$ is proportional to the net number of spins within conventional itinerant ferromagnets [64], the transition from majority to minority spins can be qualitatively expected to produce the observed reduction of slope of $T_C(\delta)$ from $\delta \sim 0.3 - 0.4$ (steep slope) to $\delta \sim 0.5 - 0.7$ (gentle slope) (details in ***SI Appendix* Section C**). Moreover, the collective picture on the observed band dependence of magnetism of $Cr_{1+\delta}Te_2$ band structure with δ provides further validity to our overall interpretation of the k-space evolution of AHE in this system.

**Summary & Impact**

In summary, complementary electronic structure and transport studies enable us to reveal the existence of a characteristic semi-metallic band region with prominent Berry curvature in the self-intercalated TMD $Cr_{1+\delta}Te_2$. With varying doping, ARPES measurements capture the rigid-like shift of the electronic band structure, realizing tunable Berry curvature at the chemical potential. Consistently, transport measurements reveal a doping-dependent sign flip in the intrinsic anomalous Hall conductivity whose *k*-space Berry curvature origin is elucidated via DFT calculations. The confluence of a prominent semi-metallic band region and flexibility in accommodating intercalated atoms render $Cr_{1+\delta}Te_2$ a unique TMD platform for tailoring $k$-space Berry curvature.

Our findings also point $Cr_{1+\delta}Te_2$ as a unique model system for investigating exotic phenomena intertwining Berry curvature physics between real and momentum spaces [35,36]. The presence of a large $k$-space Berry curvature at the chemical potential is expected to lend itself to substantial tunability and redistribution, even in the presence of magnetic perturbations corresponding to moderate energy scales [19,20,21]. Particularly intriguing is the region with δ ~ 0.5 and above (**Fig. 6C**), where the interplay of myriad magnetotransport phenomena may result in the deviation of $\sigma_{xy}^{A_{sat}}$ vs. $\sigma_{xx}^2$ from linearity (**Fig. 5A**). Interestingly, this critical doping hosts change of magnetic anisotropy from out-of-plane to in-plane [53]. One possible cause for these intriguing effects may be the enhanced magnetic fluctuation/frustration, which may manifest itself in additional contributions to Hall transport measurements. It is interesting to note that recent study also point magnetic fluctuation/frustration driven enhancement of longitudinal thermopower around this critical doping level δ~0.5 [55]. Our work opens the door for probing and tailoring the interplay of rich, intertwined magnetic effects in real- and momentum-space under controlled external magnetic field for future experimental and theoretical studies in $Cr_{1+\delta}Te_2$.

## Materials and Methods

### Film Growth and Structural Characterization

The epitaxial $Cr_{1+\delta}Te_2$ films used in this work were grown on $Al_2O_3$ (001) substrate. As described in our previous work [53], the growth was performed with an MBE system, via a two step-procedure involving film deposition, followed by in situ post deposition annealing. The annealing temperature determined the resulting composition (i.e., δ) of the $Cr_{1+\delta}Te_2$ films, which can be varied over $0.33 < \delta < 0.8$ [53]. Following detailed characterization studies performed in our previous work, the film thickness was measured using atomic force microscopy, crystallinity was confirmed by X-ray diffraction (XRD) and cross-sectional scanning transmission electron microscopy (STEM), and doping δ was determined by combining EDS with XRD lattice constants [53]. The thickness of all samples used in this study was approximately 100 nm.

### Transport and Magnetometry Measurements

Resistivity and Hall transport measurements were performed with a Quantum Design PPMS® DynaCool system. Both the longitudinal and transverse resistivity were measured simultaneously in the presence of a magnetic field to obtain the anomalous Hall effect results reported in this work. Meanwhile, magnetization measurements were performed using a Quantum Design MPMS® 3 system.

### In Situ Photoemission Spectroscopy (ARPES)

ARPES measurements were conducted on samples transferred in pristine condition from the adjacent, UHV-connected MBE system, i.e., without any ambient exposure or capping layer. The DA30HL (Scienta Omicron™) electron analyzer was used to study the photoexcited electrons with instrumental energy resolution < 10 meV. All ARPES data shown in this report were obtained at 15 K, using He I light source ($h\nu = 21.2$ eV), which provides higher energy resolution and signal-to-noise ratio comparing to He II light source ($h\nu = 40.8$ eV). We note that observe qualitatively similar Fermi pockets were also seen, using He II light source ($h\nu = 40.8$ eV) regardless of much lower intensity (see **SI appendix** Fig. S10). In this work, the photon energy-dependence of $k_z$ is not expected to significantly affect the key results, namely the observation of semi-metallic band at the M-point and the rigid-like shift of all near-$E_F$ bands with doping, δ. The weaker spectral intensity around $\overline{M}$ relative to that around $\overline{\Gamma}$ likely arises from orbital dependent cross-section, since at ~21 eV, the photoionization cross-section for *d* orbitals (near $\overline{M}$) is weaker than that of *p*-orbitals (near $\overline{\Gamma}$) [65].

### DFT Band Calculations

First principles DFT calculations were performed using an established framework [66] utilizing the generalized gradient approximation (GGA) in the Perdew-Burke-Ernzerhof (PBE) form [67]. Projector-augmented-wave [68] wave functions with energy cut-off of 500 eV were used within the Vienna ab-initio simulation package (VASP) [69,70]. The structures were optimized until the residual forces were less than $10^{-3}$ eV/Å, and the self-consistency criteria for convergence was set to $10^{-6}$ eV. Γ-centered Monkhorst-Pack [71] grids of size $12 \times 12 \times 8$, and on-site

Coulomb repulsion energy, U = 2 eV [54] were used for CrTe$_2$, and the band structure was obtained from the PBE calculations. In these calculations, the spin orientation is fully aligned along the out-of-plane (OP) direction, consistent with the intended comparison with anomalous Hall signal from the fully saturated, OP magnetic state.

**Anomalous Hall Conductivity Calculations**

For simplicity, we focus on $\sigma_{xy}{}^{\text{calc}}(E)$ from the parent CrTe$_2$ whose comparison with $\sigma_{xy}{}^{\text{expt}}(\delta)$ is justified given the consistency of band structures and observed rigid band shift with doping (see main text for details). A Hamiltonian derived from maximally localized Wannier functions was obtained using the WANNIER90 package [72] and the anomalous Hall conductivity was calculated using WannierTools [73]. The following expressions were used to calculate the Berry curvature ($\Omega_{n,xy}(k)$), total Berry curvature ($\Omega_{xy}(E,k)$) and anomalous Hall conductivity ($\sigma_{xy}^{\text{calc}}$) respectively from the Hamiltonian derived from maximally localized Wannier functions [74].

$$\Omega_{n,xy}(k) = -2\,\text{Im} \sum_{m \neq n} \frac{\langle u_n | \frac{\partial \widehat{H}}{\partial k_x} | u_m \rangle \langle u_m | \frac{\partial \widehat{H}}{\partial k_y} | u_n \rangle}{[\epsilon_m(k) - \epsilon_n(k)]^2} \qquad (2)$$

$$\Omega_{xy}(E,k) = \sum_n f_n(E)\Omega_{n,xy}(k) \qquad (3)$$

$$\sigma_{xy}^{\text{calc}}(E) = -\frac{e^2}{\hbar} \int_{BZ} \frac{dk}{(2\pi)^3} \Omega_{xy}(E,k) \qquad (4)$$

, where the $f_n(E)$ is the Fermi-Dirac distribution and $E$ is the chemical potential.

**Acknowledgments**

Y. O. acknowledges support from CREST under Grants No. JPMJCR1812. A.S. and K.H.K. acknowledge the support of the SpOT-LITE program (Grant No. A18A6b0057), funded by Singapore's RIE2020 initiatives. FCC and CHH acknowledge support from the National Center for Theoretical Sciences and the Ministry of Science and Technology of Taiwan under Grants No. MOST-110-2112-M-110-013-MY3. FCC and CHH also acknowledge the National Center for High-performance Computing for computer time and facilities.

**Figures and Table**

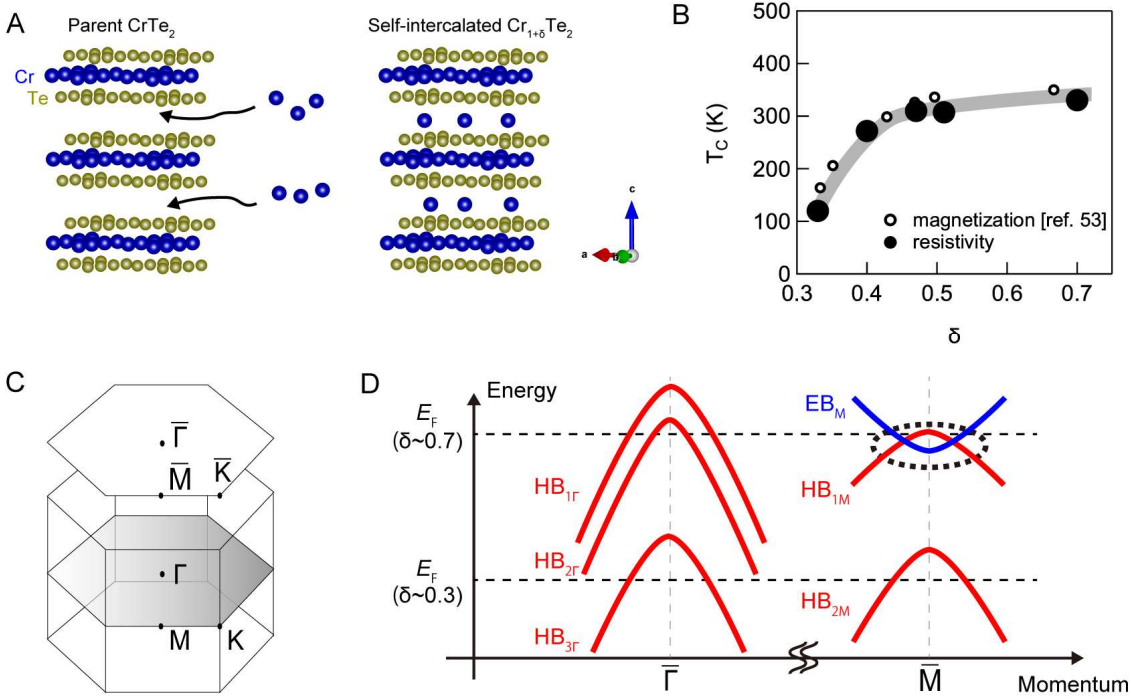

**Fig. 1.** Atomic and Electronic Structure of Self-Intercalated $Cr_{1+\delta}Te_2$. (A) Schematic crystal structure of the parent TMD $CrTe_2$ (left) and its self-intercalated counterpart $Cr_{1+\delta}Te_2$ (right). The $Cr_\delta$ atoms are intercalated between weakly van der Waals coupled Te layers (B) Doping dependence of Curie temperature, $T_c$, obtained from kink in resistivity measurements in this work (filled symbol), and compared with our previous results from magnetometry measurements (empty symbol) [53]. Grey line is a guide-to-the-eye. (C) Schematics of surface and bulk Brillouin zones for $Cr_{1+\delta}Te_2$, with key high-symmetry points indicated. (D) Schematic near Fermi energy ($E_F$) band structure of $Cr_{1+\delta}Te_2$ (details in Fig. 3, 4, 6) showing three bands each around $\bar{\Gamma}$ and $\bar{M}$, labeled as HB (hole-like) or EB (electron-like) respectively. Dotted circle highlights the semi-metallic band region central to this work, while dashed lines approximate the $E_F$ for the lower and upper values of δ realized here (see text for details).

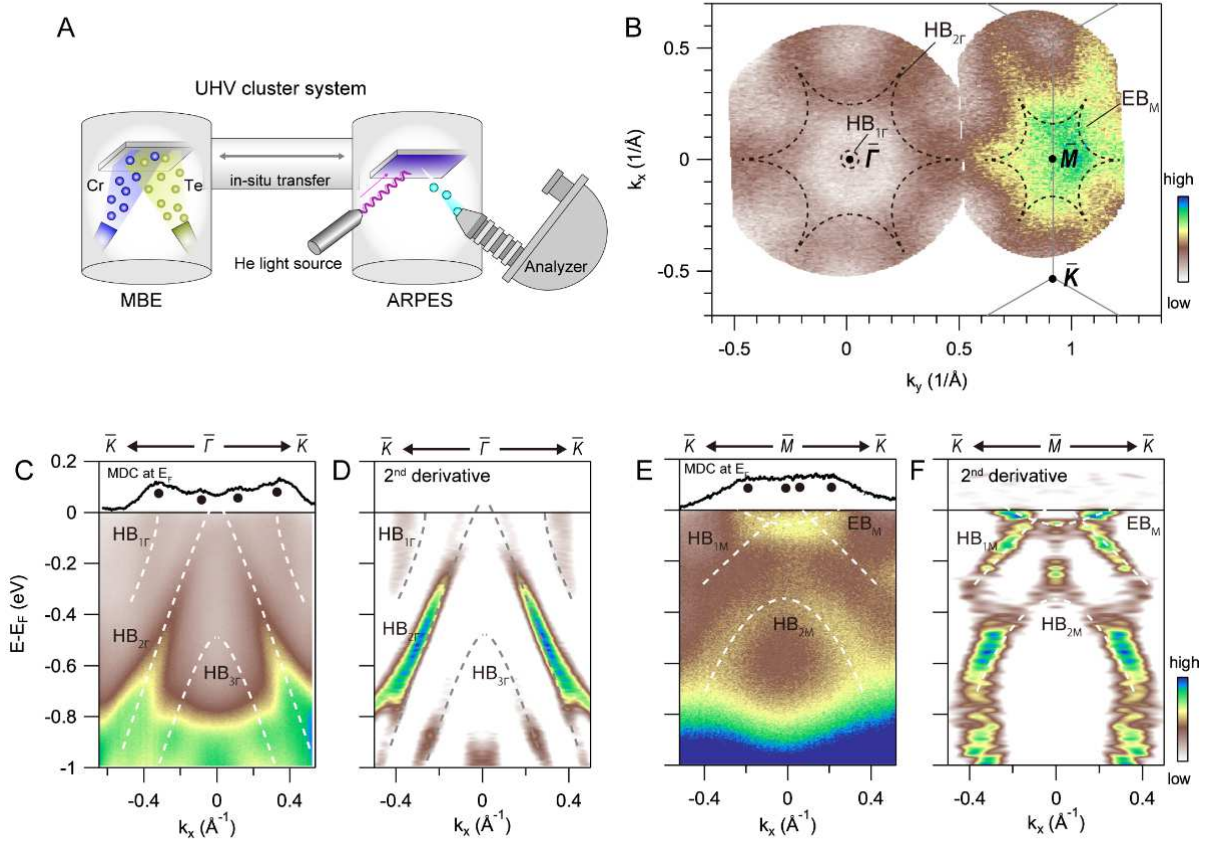

**Fig. 2.** In-situ angle resolved photoemission spectroscopy (ARPES) of $Cr_{1+\delta}Te_2$ films ($\delta$=0.6) at 15 K, with He-I light source (h$\nu$ = 21.2 eV). (A) Schematic of the ultra-high vacuum (UHV) cluster system used in this study, equipped with molecular beam epitaxy (MBE) growth system and ARPES using He based discharge lamp. (B) ARPES mapping of near-$E_F$ spectral weight for $\delta$ = 0.6 sample, representative of the Fermi surface, obtained by integrating between 0 meV and -75 meV from $E_F$. Black dashed lines are guide to the eyes. c-f Measured band dispersion for $\delta$= 0.6 along high-symmetry directions (C)-(D) $\overline{K} - \overline{\Gamma} - \overline{K}$ and (E)-(F) $\overline{K} - \overline{M} - \overline{K}$, respectively (He-I source). Here, (C, E) show the spectral intensity and (D, F) its second derivative respectively. Overlaid dashed white/grey lines bands indicate the identifiable bands (c.f. Fig. 1(D)). Insets to (C, E) show respective momentum distribution curves (MDCs) at $E_F$, with black dots indicating band crossings.

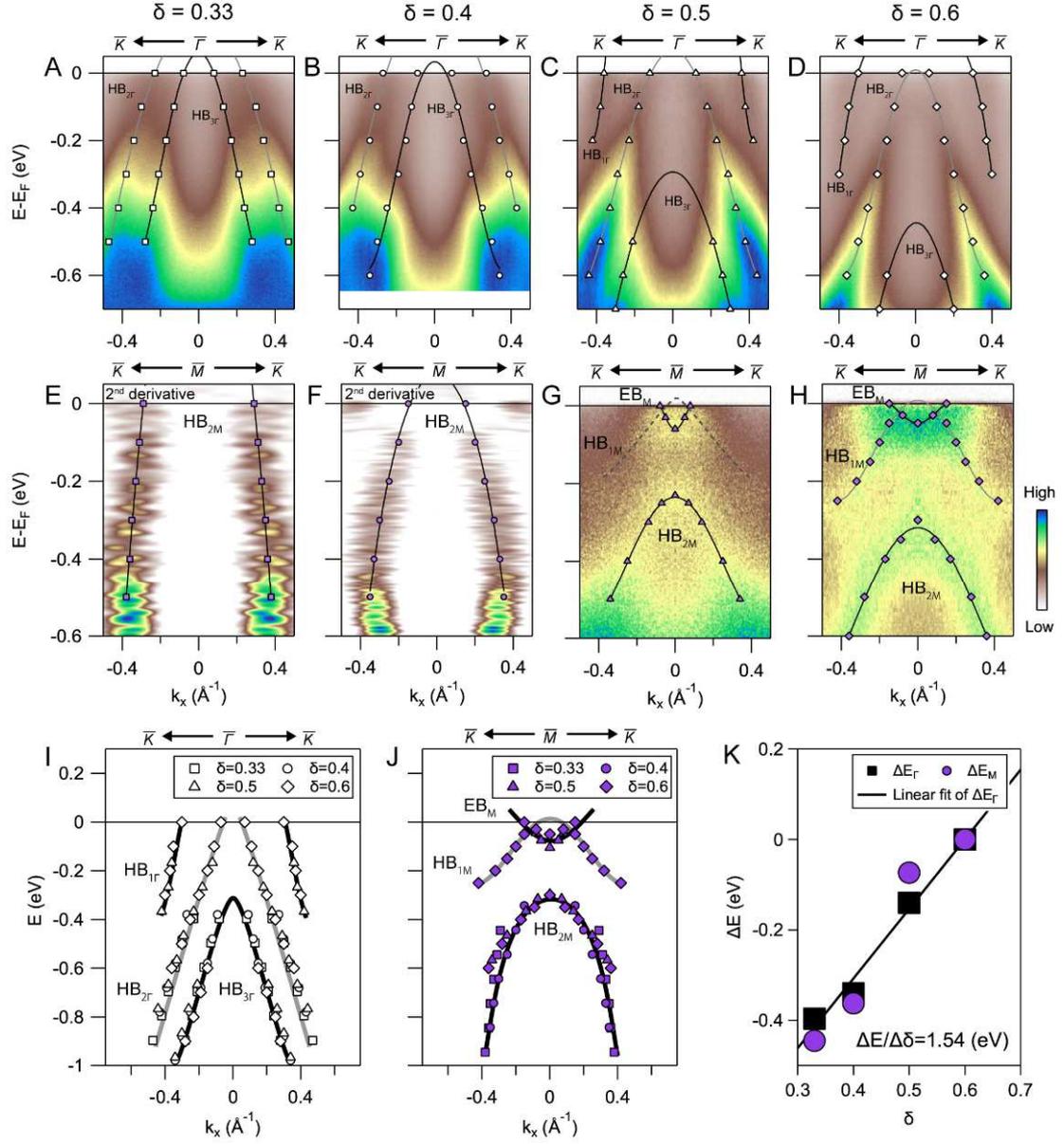

**Fig. 3.** Doping Evolution of Measured band structure at 15 K with He-I light source (h$\nu$ = 21.2 eV). (A-H) Evolution of near-$E_F$ band dispersion along (A-D) $\overline{K} - \overline{\Gamma} - \overline{K}$ and (E-H) $\overline{K} - \overline{M} - \overline{K}$ (E, F show second derivative) for samples with δ=0.33, 0.4, 0.5 and 0.6 respectively (left to right). Overlaid markers denote fits to the bands, used in (I-J) for comparisons. Black and gray dashed lines are guide to the eyes. (I,J) Measured dispersions of the respective bands across all δ at $\overline{K} - \overline{\Gamma} - \overline{K}$ (I) and $\overline{K} - \overline{M} - \overline{K}$ (J), collapsed onto single curves by shifting all data points for each δ by a constant energy ($\Delta E_\Gamma(\delta)$ and $\Delta E_M(\delta)$, details in text). Black and gray lines in (I, J) are guide to the eyes. (K) The variation of thus defined energy shifts $\Delta E_\Gamma$ and $\Delta E_M$ with δ, with the overlaid line showing a linear fit (details in text).

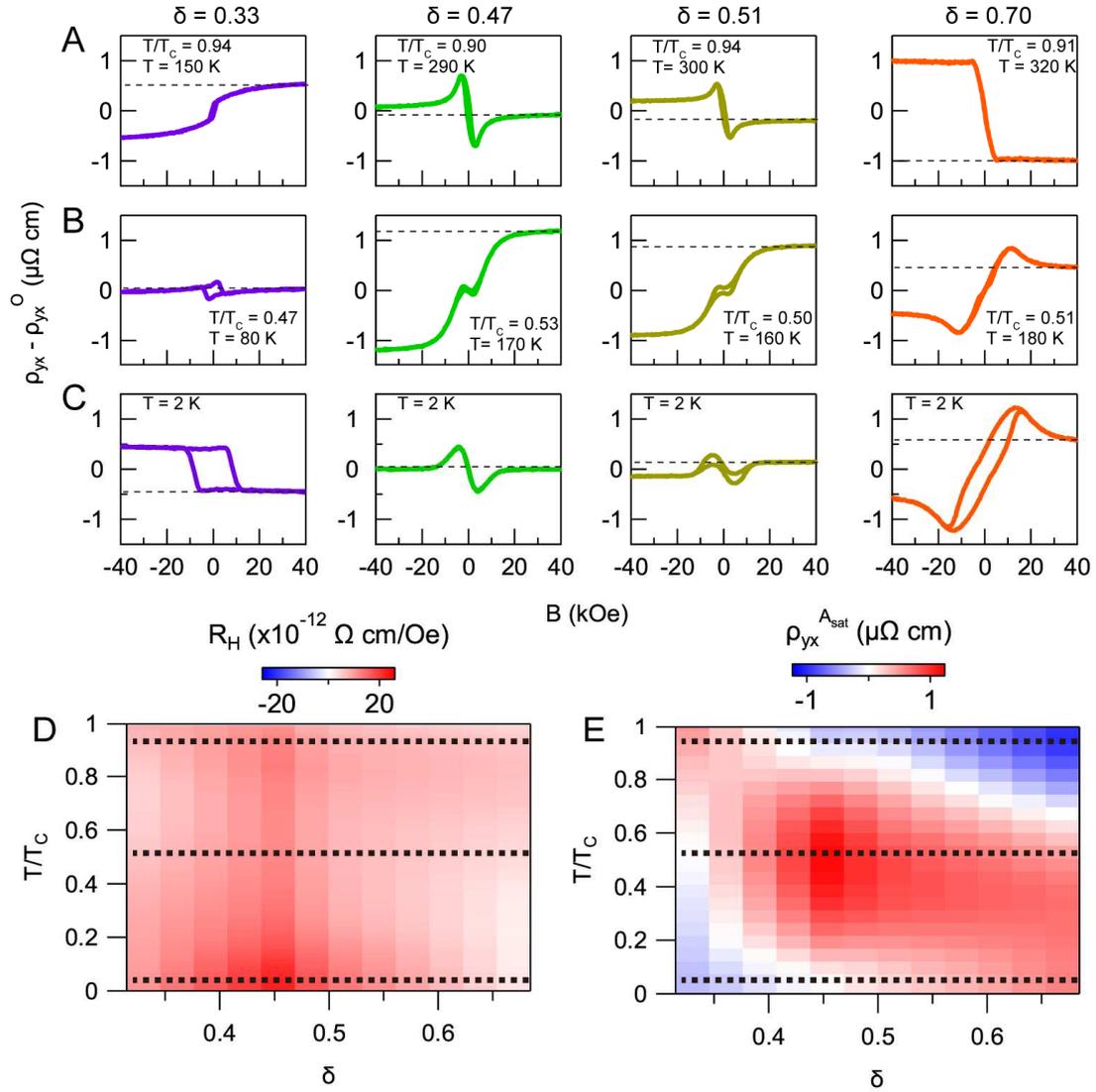

**Fig. 4.** Measured Evolution of Hall Effect. (A-C) The variation of Hall resistivity, $\rho_{yx}(B) - \rho_{yx}^O(B)$, with temperature $-T/T_c \sim 0.9$, 0.5 and $T = 2$ K (top to bottom), across samples with δ = 0.33, 0.47, 0.51 and 0.70 (left to right). Dashed lines in each panel indicated the saturated anomalous Hall resistivity, $\rho_{yx}^{A_{sat}}$. (D-E) Color plots of the Hall coefficient, $R_H$ (D), and the saturated AHE resistivity, $\rho_{yx}^{A_{sat}}$ (E) as a function of $T/T_c$ and δ. $R_H$ is deduced from a linear fit at high fields, while $\rho_{yx}^{A_{sat}}$ is obtained from saturated magnitude of $\rho_{yx}(B) - \rho_{yx}^O(B)$ (e.g., dotted lines in A-C).

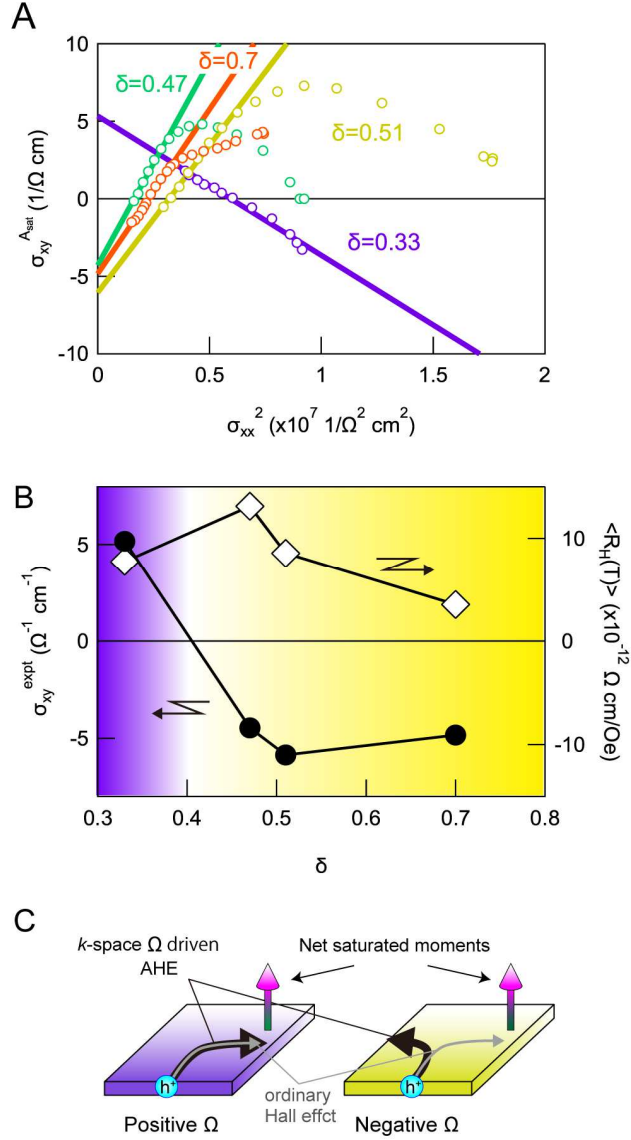

**Fig. 5.** Scaling Analysis of Anomalous Hall Effect. (A) Plots of $\sigma_{xy}^{A_{sat}}$ vs. $\sigma_{xx}^2$ for samples with δ = 0.33, 0.47, 0.51 and 0.7 (data shown for $T/T_c < 0.9$). For each case, overlaid lines indicate the linear (i.e. $\sigma_{xy}^{A_{sat}} \propto \sigma_{xx}^2$) region (details in text). (B) Doping dependence of measured intrinsic anomalous Hall conductivity $\sigma_{xy}^{expt}$ (left axis, procedural details in text) and temperature averaged high field Hall coefficient $\langle R_H(T) \rangle$. (C) Schematics for Berry curvature effects on itinerant hole carriers, and the resulting sign change of AHE ($\sigma_{xy}^{expt}$) for magnetized samples.

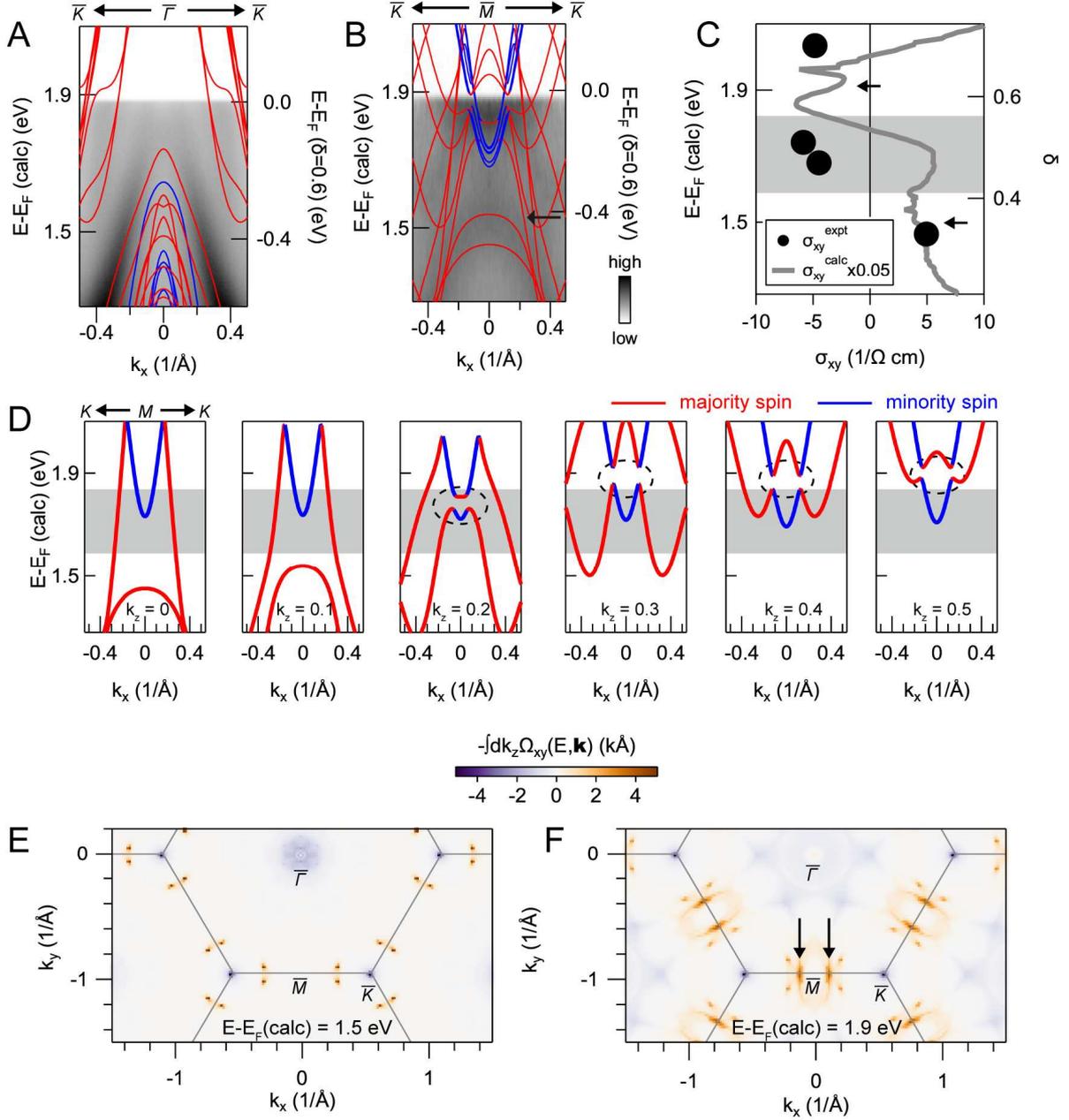

**Fig. 6.** Momentum-Space Analysis of Berry Curvature Effects. (A-B) Comparison between ARPES-measured band dispersion for $Cr_{1+\delta}Te_2$ with $\delta = 0.6$ (shown in grayscale) with DFT-calculated band structure for $CrTe_2$ (energy shifted by 1.9 eV, overlaid as lines) along (A) $\overline{K} - \overline{\Gamma} - \overline{K}$ and (B) $\overline{K} - \overline{M} - \overline{K}$. Left and right axes are E-$E_F$ for DFT and ARPES respectively. The $E_F$ for ARPES data ($\delta \sim 0.6$) is set to 1.9 eV higher than that for DFT data ($\delta=0$). DFT-calculated bands comprise major (red) and minor (blue) spin orientations, and results for varying $k_z$ are all overlaid

for comparison with ARPES data, which lack $k_z$ resolution. (C) Comparison between DFT-calculated $\sigma_{xy}^{\text{calc}}(E)$ (grey line, from Eqn. (4) in method section, details in text) and measured $\sigma_{xy}^{\text{expt}}(\delta)$ (black dots, from Fig. 5(B)). One-to-one comparison between calculations ($E$) and experiments ($\delta$) is enabled by matching the Fermi energies in (A)-(B), together with the measured $\Delta E_\text{F}/\Delta\delta$ (Fig. 3K, details in text). Horizontal arrows indicate the energies used for (E) and (F) respectively. (D) The $k_z$ resolved dispersion of the DFT calculated semi-metallic band region near EF along $\overline{K} - \overline{M} - \overline{K}$ (c.f. collated in (B)). Colors indicate spin orientation (c.f. (A)), while dashed circle highlights spin-orbit coupling (SOC) driven opening of a band gap, hosting large k-space Berry curvature. (E-F) Color plots of the distribution of calculated Berry curvature, $\int dk_z \Omega_{xy}(E, k)$, across the $k_x - k_y$ plane at (E) $E - E_F(\delta = 0) = +1.5$ eV and (F) +1.9 eV (horizontal arrows in (C)). Arrows highlight a region near $M$-point, hosting a large Berry curvature contribution within (F).

# Supplementary material for
# Widely Tunable Berry curvature in the Magnetic Semimetal $Cr_{1+\delta}Te_2$

Y. Fujisawa[1], M. Pardo-Almanza[1], C. H. Hsu[1,2,3], A. Mohamed[1], K. Yamagami[1], A. Krishnadas[1], F. C. Chuang[2,3,4], K. H. Khoo[5], J. Zang[6,7], A. Soumyanarayanan[8,9], Y. Okada[1,]

[1] Quantum Materials Science Unit, Okinawa Institute of Science and Technology (OIST), Okinawa 904-0495, Japan

[2] Department of Physics, National Sun Yat-sen University, Kaohsiung 80424, Taiwan

[3] Physics Division, National Center for Theoretical Sciences, Taipei 10617, Taiwan.

[4] Department of Physics, National Tsing Hua University, Hsinchu 30013 Taiwan.

[5] Institute of High Performance Computing, Agency for Science Technology and Research, 138632 Singapore

[6] Department of Physics and Astronomy, University of New Hampshire, Durham, NH 03824, USA.

[7] Materials Science Program, University of New Hampshire, Durham, NH 03824, USA.

[8] Department of Physics, National University of Singapore, 117551 Singapore

[9] Institute of Materials Research and Engineering, Agency for Science Technology and Research, 138634 Singapore

## Section A

### Validation of the energy shift in the calculated band dispersion

To match the DFT calculated band structure and experimental ARPES dispersion of $Cr_{1+\delta}Te_2$ ($\delta$ =0.6), we have shifted the calculated band by 1.9 eV as is shown in the main text (see also **Fig. 6A-B**). In this case, if we extrapolate the shift to $\delta$ =0 case assuming the rigid band shift, the difference from the Femi energy of calculated $CrTe_2$ would be 1.1 eV (see intercept on the left axis in **Fig. S6a**). This mismatch is presumably due to difficulty in implementing electronic correlation effect precisely in the DFT calculation. Similar difficulty has been discussed in literature [1,2].

In order to check this difference is within a realistic range, we have calculated occupied *d* electron number of Cr. **Fig. S6b** shows calculated DOS of 3d states for $CrTe_2$. Here, $N_{up}$ and $N_{down}$ denote majority spin and minority spin, respectively. In **Fig. 6b**, bottom axis is as calculated energy in DFT, without correction. We found that the difference in the *d* electron number due to 1.1 eV shifting is reasonably small (0.2 electrons with majority spin, 0 electrons with minority spin). Thus, we conclude that the 1.1 eV shift is within reasonable range.

## Section B

### Weyl points

**Fig. S7a** shows calculated band structure of $CrTe_2$ with spin-orbit coupling. The experimentally observed bands with band crossing are shown in magenta and cyan lines in the unoccupied state. By focusing on the bands, we find Weyl points by calculating the chirality. We found 48 Weyl points (24 pairs) in total as shown in **Supplementary table1**. The projection of Weyl points in $k_x - k_y$ and $k_x - k_z$ plane are shown in **Fig. S7b** and **7c**, respectively. Furthermore, the band structures of Weyl points are shown in **Fig. S8**. We can find the Weyl points labeled by the same color have same band structure. All of them are found to be located at +2.3 eV which is higher energy than our accessible energy scale (up to 1.9 eV) and around the Γ point. Thus, we think that Berry curvature from Weyl points play a minor role for controlling experimental Hall effect. On the other hand, there is no such candidate below 1.9 eV, suggesting the experimentally observed band crossing is topologically trivial.

**Section C**

**Validity of energy shift of calculation based on δ dependent $T_C$**

**Fig. S9** shows comparison between energy integrated net spin density $N_{up}$-$N_{down}$ (left axis) and Currie temperature $T_C$ (right axis). Here, $N_{up}$ and $N_{down}$ are calculated by integrating spin dependent density of state from -5 eV (see brome lines in **Fig. S6b**). As seen from this comparison, the energy (and δ) where kink in Tc is observed matches very well with the maximum in the magnetic moment. We think that this qualitative agreement supports the validity of overall comparison between experimental ARPES and calculated band structure.

For δ < 0.33, however, it is known that $T_C$ increases as approaching $CrTe_2$ with a $T_C$ of ~ 300 K [3], although it is difficult to experimentally access in this doping region by using the same method employed in this study. The possible reason for the enhancement of $T_C$ as decreasing δ might be explained as follows. When δ is first increased from 0, most of the intercalated Cr interact with Cr in other layers. The nearest neighbor interlayer interaction is usually antiferromagnetic [4], and this tends to lower $T_C$ as the mean field theory states that $T_C$ is proportional to the sum of exchange interaction strengths. However, as more Cr atoms are intercalated, it also leads to more intralayer ferromagnetic interactions within the intercalated layer and that tends to increase $T_C$. Beyond phenomenological interpretation mentioned above, we think full understanding of doping dependent $T_C$ requires further future studies.

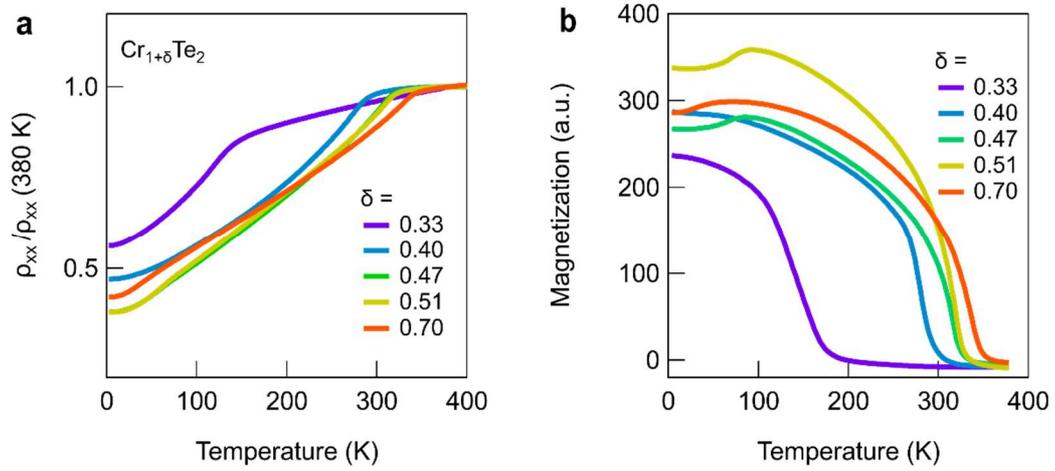

**Fig. S1.** Estimation of Curie temperature from resistivity and magnetization. a, b, Temperature dependence of normalized resistivity at zero field (a) and magnetization taken along easy axis (b). Magnetic field of 1000 Oe is applied along out-of-plane for the samples with $\delta = 0.33$ and 0.40, and along in-plane for $\delta = 0.47$, 0.51, and 0.70. $T_C$ is estimated from the temperature where kink is observed for resistivity, and from the temperature where sudden increase is observed for magnetization.

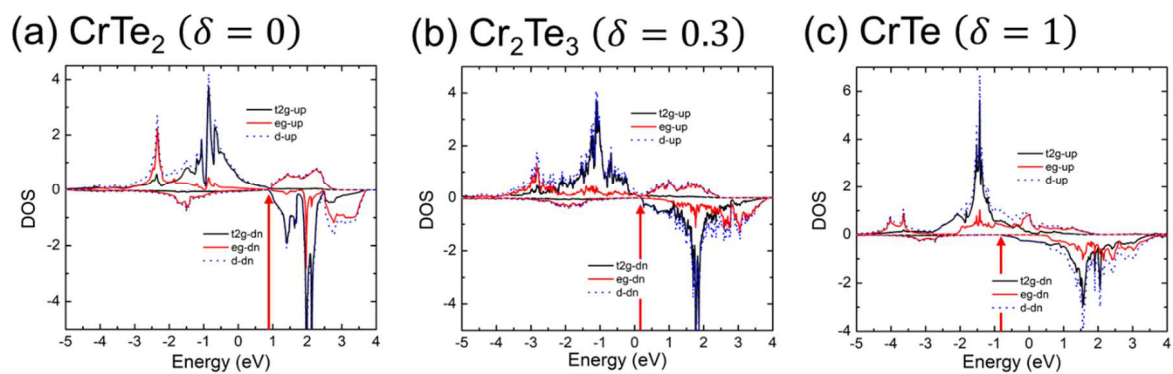

**Fig. S2.** Reproducing the trend of the rigid band shift in $Cr_{1+\delta}Te_2$ by DFT calculation. a,b,c partial density of state (DOS) of Cr in $CrTe_2$ (a), $Cr_2Te_3$ (b), CrTe (c), respectively. Qualitatively, one can see the entire DOS shift to lower energy as increasing δ, demonstrating the electron doping by δ. Quantitatively, the band bottom of minority $t_{2g}$ band moves as increasing δ from +0.8 eV ($CrTe_2$) to -0.8 eV (CrTe) as indicated by red arrows. The amount of the shift (1.6 eV) is roughly consistent with the value obtained from the experiment (see **Fig. 3k** and **Fig. S6a**).

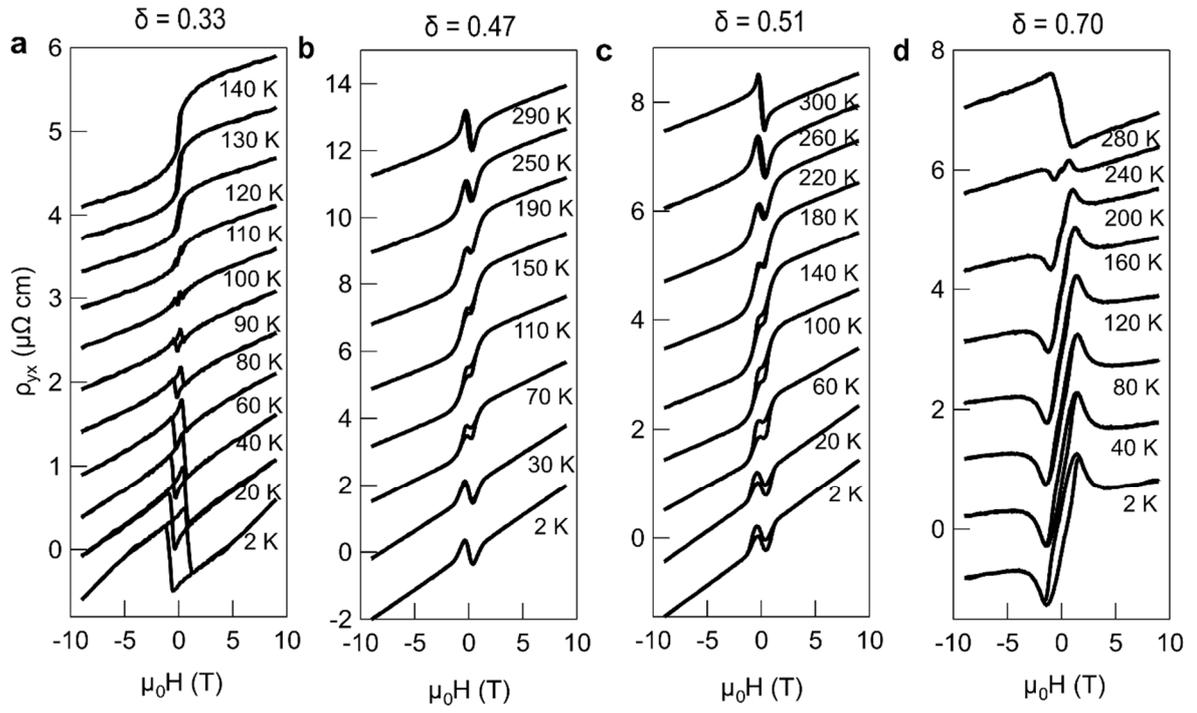

**Fig. S3.** Dataset extension for temperature and doping dependent $\rho_{yx}$. $\rho_{yx}$ for various doping levels at different temperatures.

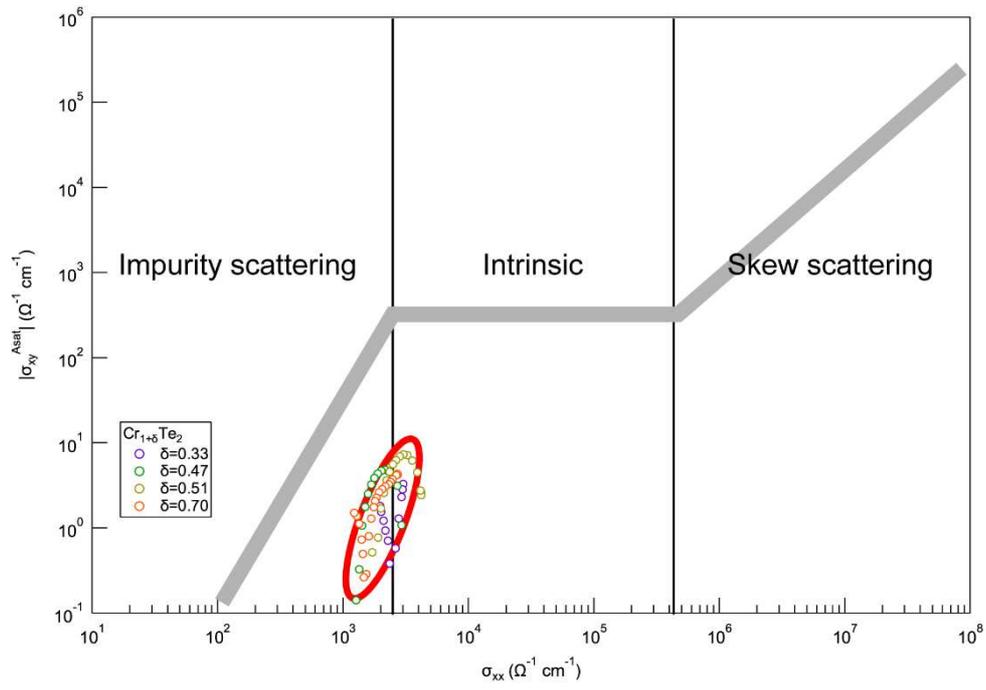

**Fig. S4.** Position of our $Cr_{1+\delta}Te_2$ films on the universal curve. Position of of our $Cr_{1+\delta}Te_2$ films on the universal curve. As described in the main text, our samples located a region where the intrinsic and extrinsic terms coexist.

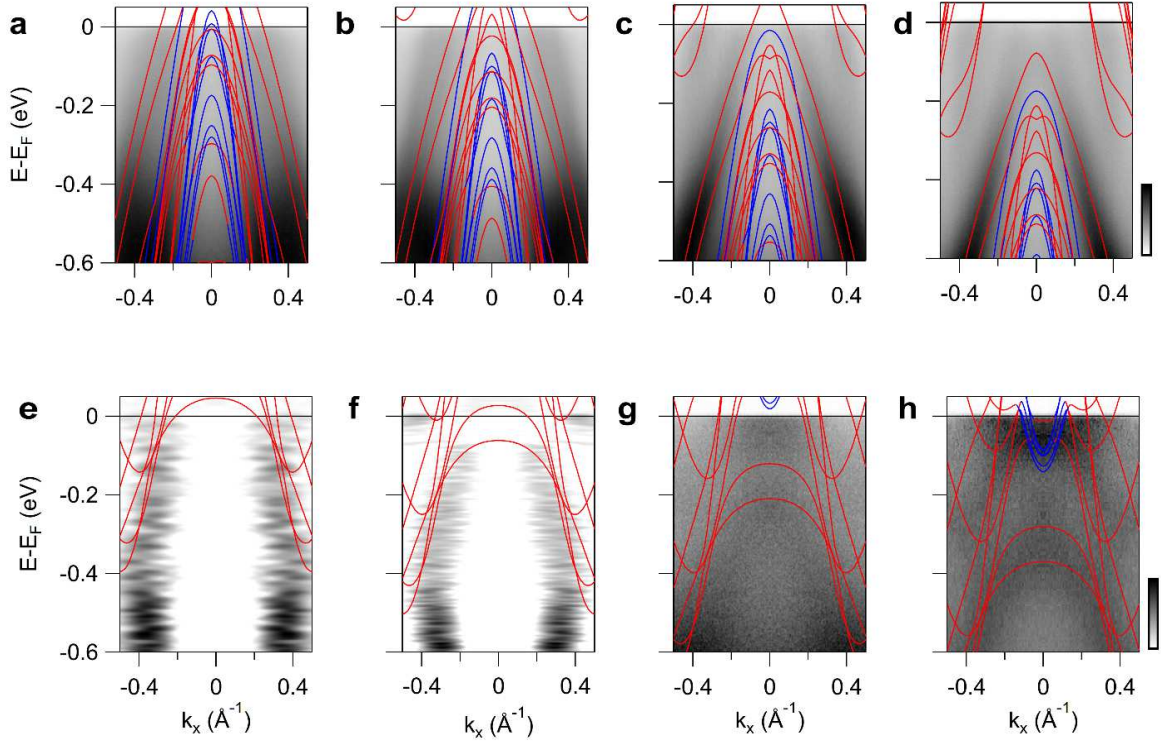

**Fig. S5.** Correspondence between ARPES data and theoretical calculation. ARPES data for δ = 0.33, 0.40, 0.50, 0.60 which are the same as **Fig. 3a-h** in the main body, but with gray scale. Red and blue lines are calculated band structures of CrTe$_2$ with spin up and down, respectively. The choice of energy shift (1.82 eV) for **d** and **h** are made so that characteristic branches of bands can be reproduced. The energy shift for the other data (**a-c**, **e-g**) is determined by considering rigid energy shift in **Fig. 3k**. Each comparison suggests good agreement to conclude that the band evolution is basically considered to be rigid band shift.

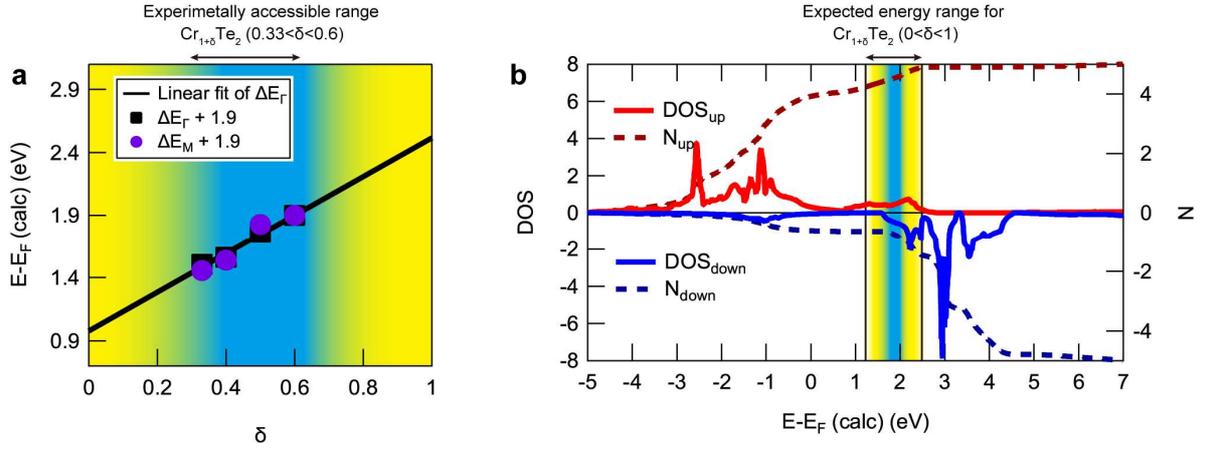

**Fig. S6.** Evolution of the calculated band structure in $Cr_{1+\delta}Te_2$ with increasing $\delta$. **a** Fermi energy for the samples in this ARPES study. The energy for $\delta=0.6$ is determined by matching DFT calculated band and ARPES dispersion as shown in **Fig. 6a** and **b**. The energy for the other samples are determined from the value of 1.9 eV for $\delta=0.6$ and experimentally determined energy slope ($\Delta E_F/\Delta\delta$) as shown in **Fig. 3k**. The black line is obtained by linear fitting of $\Delta E_\Gamma$. **b** The density of state (DOS) for majority and minority spins and their integrations (electron number, N) as a function of energy. Highlighted area with blue indicates experimentally accessible range, while the one with yellow indicates expected energy range from $CrTe_2$ to $CrTe$ assuming rigid band shift.

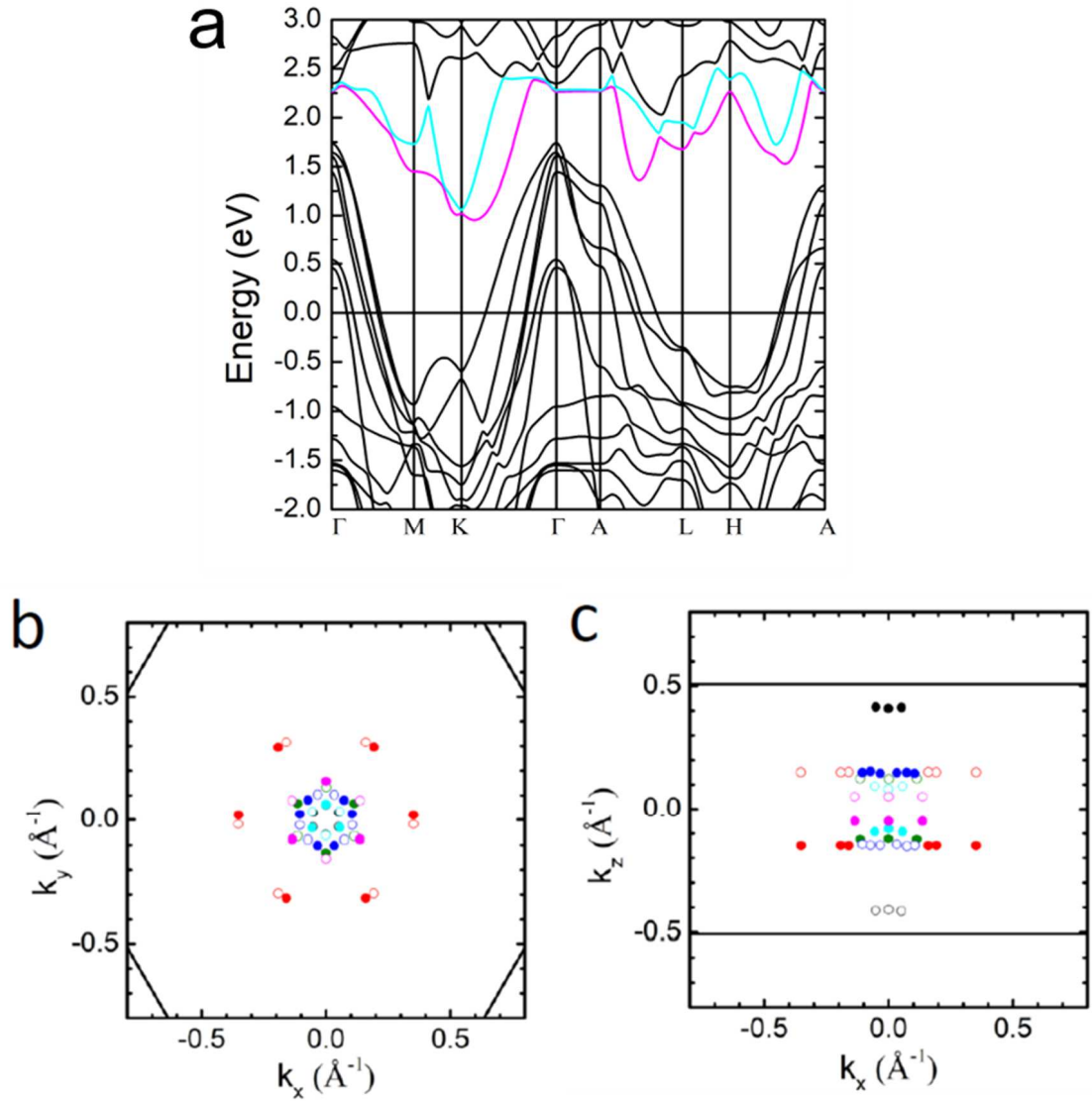

**Fig. S7.** Weyl points in CrTe$_2$. **a,** Band structure of CrTe$_2$ in the ferromagnetic state. The bands shown in magenta and cyan color denote where we search for the Weyl points. The projection of all Weyl points in **b,** $k_x - k_y$ and **c,** $k_x - k_z$ plane. The black line is the surface Brillouin zone. Different colors correspond to the color in **supplementary table 1** and represent a set of Weyl points with the same band structures. The solid and empty circles denote the chirality of $+1$ and $-1$, respectively.

|   | $k_1$ | $k_2$ | $k_3$ | Energy | Chirality |
|---|---|---|---|---|---|
| $WP_1^+$ | 0.000 | -0.061 | 0.410 | 2.304 | +1 |
| $WP_1^-$ | 0.000 | 0.061 | -0.410 | 2.304 | -1 |
| $WP_2^+$ | -0.051 | 0.029 | 0.415 | 2.303 | +1 |
| $WP_2^-$ | 0.051 | -0.029 | -0.415 | 2.303 | -1 |
| $WP_3^+$ | 0.001 | 0.053 | -0.442 | 2.304 | +1 |
| $WP_3^-$ | -0.001 | -0.053 | 0.442 | 2.304 | -1 |
| $WP_4^+$ | 0.160 | -0.315 | -0.151 | 2.350 | +1 |
| $WP_4^-$ | -0.160 | 0.315 | 0.151 | 2.350 | -1 |
| $WP_5^+$ | -0.352 | 0.019 | -0.150 | 2.351 | +1 |
| $WP_5^-$ | 0.352 | -0.019 | 0.150 | 2.351 | -1 |
| $WP_6^+$ | 0.352 | 0.019 | -0.150 | 2.350 | +1 |
| $WP_6^-$ | -0.352 | -0.019 | 0.150 | 2.350 | -1 |
| $WP_7^+$ | 0.193 | 0.295 | -0.150 | 2.351 | +1 |
| $WP_7^-$ | -0.193 | -0.295 | 0.150 | 2.351 | -1 |
| $WP_8^+$ | -0.160 | -0.315 | -0.151 | 2.350 | +1 |
| $WP_8^-$ | 0.160 | 0.315 | 0.151 | 2.350 | -1 |
| $WP_9^+$ | -0.193 | 0.295 | -0.150 | 2.351 | +1 |
| $WP_9^-$ | 0.193 | -0.295 | 0.150 | 2.351 | -1 |
| $WP_{10}^+$ | 0.105 | 0.023 | 0.145 | 2.357 | +1 |
| $WP_{10}^-$ | -0.105 | -0.023 | -0.145 | 2.357 | -1 |
| $WP_{11}^+$ | 0.033 | -0.103 | 0.149 | 2.357 | +1 |
| $WP_{11}^-$ | -0.033 | 0.103 | -0.149 | 2.357 | -1 |
| $WP_{12}^+$ | -0.106 | 0.022 | 0.149 | 2.357 | +1 |
| $WP_{12}^-$ | 0.106 | -0.022 | -0.149 | 2.357 | -1 |
| $WP_{13}^+$ | 0.072 | 0.080 | 0.149 | 2.357 | +1 |
| $WP_{13}^-$ | -0.072 | -0.080 | -0.149 | 2.357 | -1 |
| $WP_{14}^+$ | -0.073 | 0.081 | 0.153 | 2.357 | +1 |
| $WP_{14}^-$ | 0.073 | -0.081 | -0.153 | 2.357 | -1 |
| $WP_{15}^+$ | -0.033 | -0.102 | 0.145 | 2.357 | +1 |
| $WP_{15}^-$ | 0.033 | 0.102 | -0.145 | 2.357 | -1 |
| $WP_{16}^+$ | -0.001 | -0.132 | -0.123 | 2.307 | +1 |
| $WP_{16}^-$ | 0.001 | 0.132 | 0.123 | 2.307 | -1 |
| $WP_{17}^+$ | -0.114 | 0.066 | -0.125 | 2.306 | +1 |
| $WP_{17}^-$ | 0.114 | -0.066 | 0.125 | 2.306 | -1 |
| $WP_{18}^+$ | 0.113 | 0.066 | -0.125 | 2.307 | +1 |
| $WP_{18}^-$ | -0.113 | -0.066 | 0.125 | 2.307 | -1 |
| $WP_{19}^+$ | -0.056 | -0.034 | -0.094 | 2.318 | +1 |
| $WP_{19}^-$ | 0.056 | 0.034 | 0.094 | 2.318 | -1 |
| $WP_{20}^+$ | 0.056 | -0.032 | -0.093 | 2.317 | +1 |
| $WP_{20}^-$ | -0.056 | 0.032 | 0.093 | 2.317 | -1 |
| $WP_{21}^+$ | 0.001 | 0.062 | -0.082 | 2.315 | +1 |
| $WP_{21}^-$ | -0.001 | -0.062 | 0.082 | 2.315 | -1 |
| $WP_{22}^+$ | 0.136 | -0.079 | -0.050 | 2.328 | +1 |
| $WP_{22}^-$ | -0.136 | 0.079 | 0.050 | 2.328 | -1 |
| $WP_{23}^+$ | 0.000 | 0.157 | -0.050 | 2.328 | +1 |
| $WP_{23}^-$ | 0.000 | -0.157 | 0.050 | 2.328 | -1 |
| $WP_{24}^+$ | -0.136 | -0.078 | -0.050 | 2.328 | +1 |
| $WP_{24}^-$ | 0.136 | 0.078 | 0.050 | 2.328 | -1 |

**Supplementary table 1.** The list of Weyl points in CrTe$_2$. The fractional coordinates of all Weyl points in momentum space, Chirality, and the energy relative to $E_F$ are shown. Different color corresponds to the color in **Fig. S7b**.

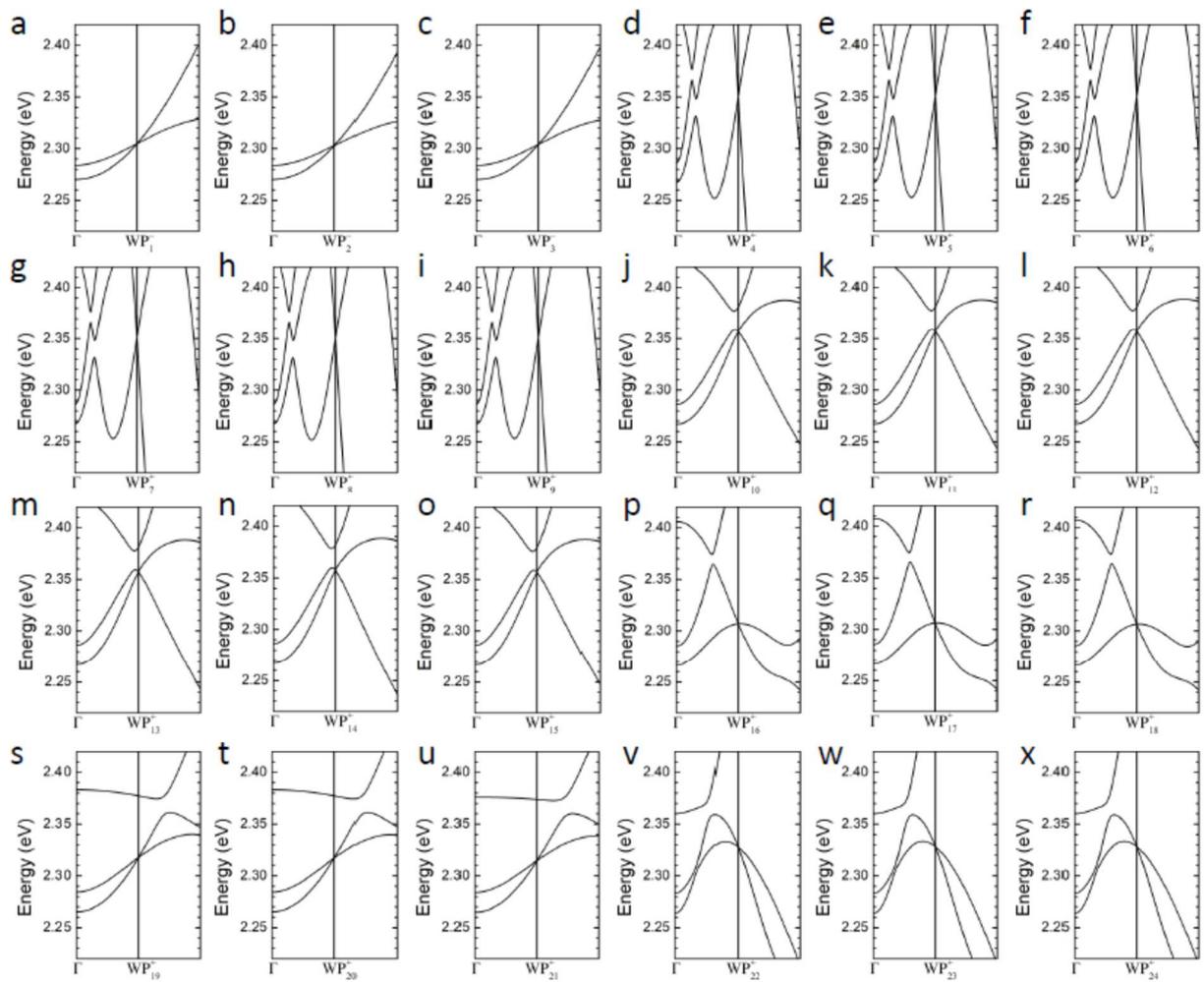

**Fig. S8.** Band structure of all Weyl points at the +2.3 eV in CrTe$_2$

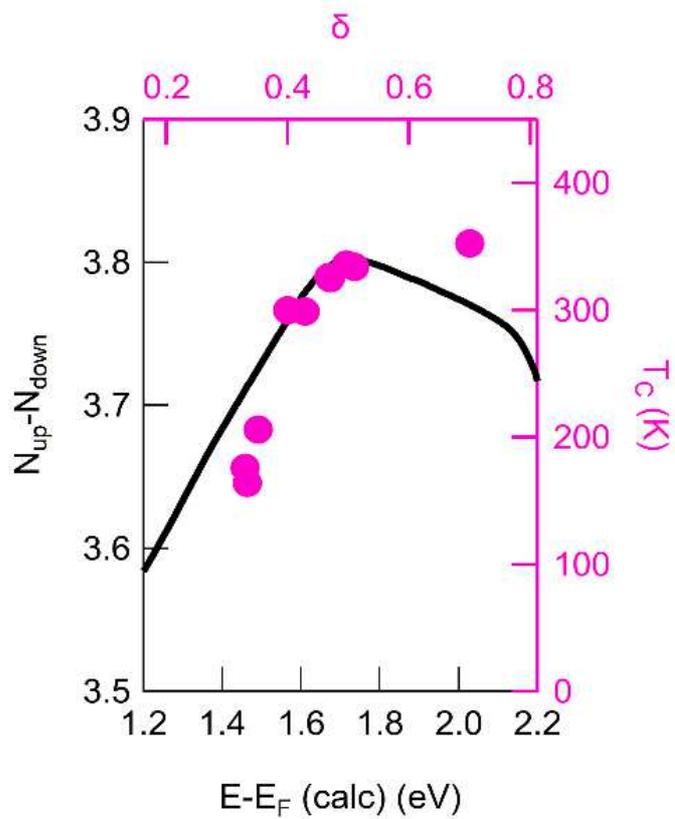

**Fig. S9.** Comparison between $T_C$ and total number of net spins ($=N_{up}-N_{down}$). $N_{up}$ and $N_{down}$ are from **Fig. S6b.**

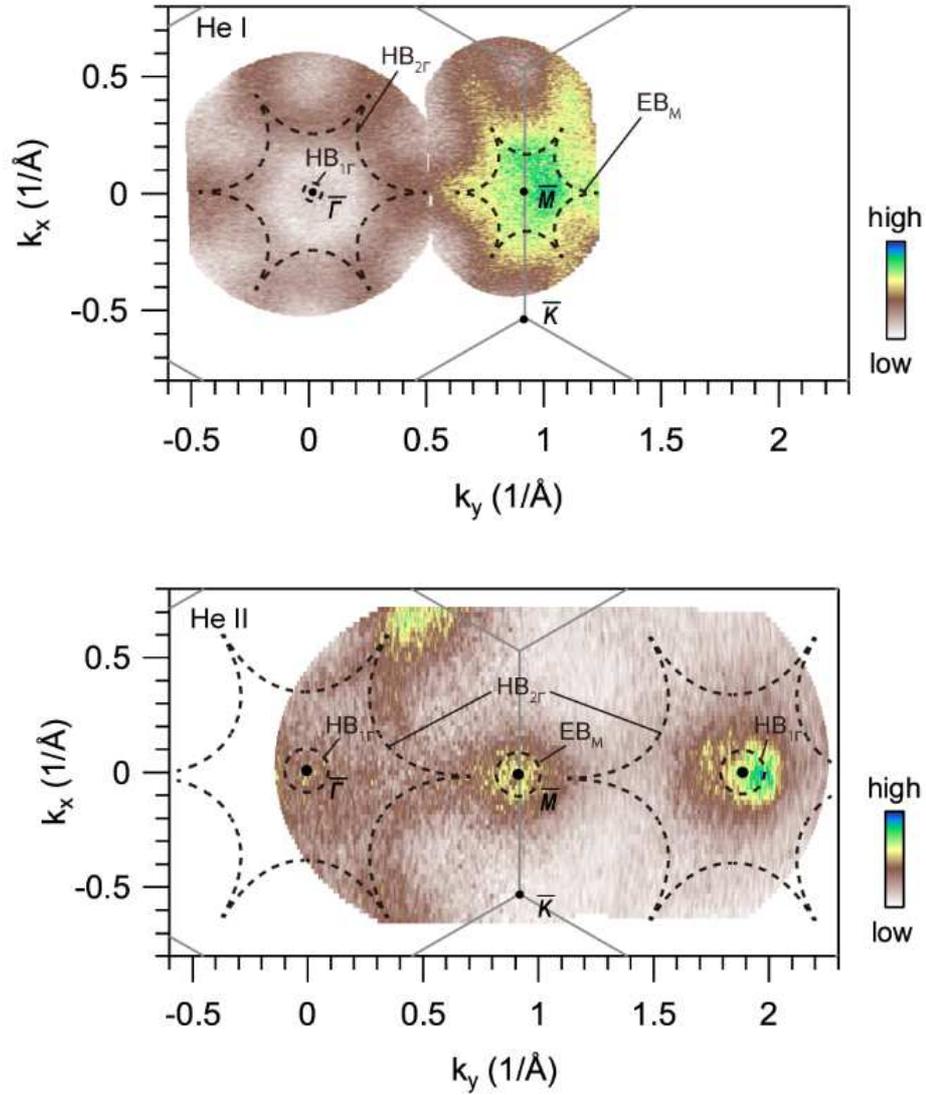

**Fig. S10.** ARPES mapping of near-$E_F$ spectral weight for δ = 0.6 sample, representative of the Fermi surface, obtained by integrating between 0 meV and -75 meV from $E_F$ taken with He I (upper) and He II (lower), respectively. Black dashed lines are guide to the eyes (see **Fig. 1D** for the definition of multiple bands).